\documentclass[twocolumn,tighten,times]{aastex63}
\usepackage{graphicx}

\newcommand{\src}{RCW 89}
\newcommand{\cha}{{\sl Chandra}}
\catcode`\@=11
\newcommand{\gapprox}{\mathrel{\mathpalette\@versim>}}
\newcommand{\lapprox}{\mathrel{\mathpalette\@versim<}}
\newcommand{\propapprox}{\mathrel{\mathpalette\@versim\propto}}
\newcommand{\@versim}[2]
  {\lower3.1truept\vbox{\baselineskip0pt\lineskip0.5truept
\ialign{$\m@th#1\hfil##\hfil$\crcr#2\crcr\sim\crcr}}}
\catcode`\@=12

\received{2020 April 10}
\revised{2020 May 6}
\accepted{2020 May 11}
\journalinfo{The Astrophysical Journal Letters}

\shorttitle{Fast Blast Wave and Ejecta in RCW 89}

\begin{document}

\title{Fast Blast Wave and Ejecta in the Young Core-Collapse Supernova Remnant MSH 15-5{\it 2}/RCW 89}

%\correspondingauthor{Kazimierz J. Borkowski}
%\email{kborkow@ncsu.edu}

\author{Kazimierz J. Borkowski}
\affiliation{Department of Physics, North Carolina State University, 
Raleigh, NC 27695-8202, USA}

\author[0000-0002-5365-5444]{Stephen P. Reynolds}
\affiliation{Department of Physics, North Carolina State University, 
Raleigh, NC 27695-8202, USA}

\author{William Miltich}
\affiliation{Department of Physics, North Carolina State University, 
Raleigh, NC 27695-8202, USA}

\begin{abstract}
  One of the youngest known remnants of a core-collapse supernova (SN) in
  our Galaxy is \object{G320.4$-$1.2}/MSH 15-5{\it 2}, containing an energetic
  pulsar with a very short (1700 yr) spindown age and likely produced
  by a stripped-envelope SN Ibc. Bright X-ray and radio
  emission north of the pulsar overlaps with an H$\alpha$ nebula \src .
  The bright X-rays there have a highly unusual and quite puzzling morphology,
  consisting of both very
  compact thermally emitting knots and much more diffuse emission of
  nonthermal origin. We report new X-ray observations of \src\ in 2017
  and 2018 with {\sl Chandra} that allowed us to measure the motions of many knots
  and filaments on decade-long time baselines.
  We identify a fast blast wave with a velocity of
  $(4000 \pm 500)d_{5.2}$ km s$^{-1}$ ($d_{5.2}$ is the distance in units of
  5.2 kpc) with a purely nonthermal spectrum, and without any radio counterpart. 
  Many compact X-ray emission knots are  
  moving vary fast, with velocities as high as 5000 km s$^{-1}$,
  predominantly radially away from the pulsar. Their spectra show that they
  are Ne- and Mg-rich heavy-element SN ejecta. They have been
  significantly decelerated upon their recent impact with the dense ambient
  medium north of the pulsar. We see 
  fast evolution in brightness and morphology of knots in just
  a few years. Ejecta knots in \src\ resemble those seen in Cas A at
  optical wavelengths in terms of their initial velocities and densities.
  They might have the same origin, still not 
  understood but presumably related to stripped-envelope SN
  explosions themselves.
  
\end{abstract}


\section{Introduction}
\label{intro}

A significant fraction of massive stars shed part or all of their
hydrogen envelope before explosion \citep[$\sim 30$\% of core-collapse 
supernovae (SNe) are of Type IIb, Ib, or Ic; e.g.][]{shivvers17}.
The mechanisms of mass loss are not
well understood.
The supernova remnant (SNR) 
Cassiopeia A is now generally thought to have resulted from a SN IIb,
based on light-echo observations \citep{krause08,rest11}.  Cas A
exhibits the presupernova wind as quasi-stationary ``flocculi'' rich
in N, but also high-velocity knots of O-rich ejecta.  The dense
ejecta knots may hold clues to instabilities in the explosion itself.
Older objects that probably resulted from stripped-envelope events
include the complex remnant MSH 15$-5{\it 2}$/RCW 89, the subject of
this Letter.

Our target was discovered as a large ($ \sim 32'$ diameter) radio
source \citep[MSH 15$-5{\it 2}$;][]{mills61}, consisting of a fairly
bright northern region and fainter southern region \citep{caswell81},
the northern one coincident with a filamentary H$\alpha$ nebula, RCW
89 \citep{rodgers60}.  The source contains one of the most extreme
rotation-powered pulsars, with a surface field of $1.5 \times 10^{13}$
G and the third-shortest known spindown timescale (about 1700 yr): PSR
B1509-58 \citep[see data in][]{kargaltsev13}, which possesses a
peculiar, anisotropic pulsar-wind nebula (PWN) in X-rays, though
radio observations \citep{gaensler99} showed no clear
counterpart.  For a distance of 5.2 kpc \citep{gaensler02}, the
distance from the pulsar to the northern optical extent of RCW
89 is about 23 pc.  At that distance, the required expansion velocity
is about 13,000 km s$^{-1}$, though optical knots have
negligible expansion speeds \citep{vandenbergh84}.

\begin{figure*}[ht!]
\begin{center}
%\epsscale{1.0}
%\plotone{f1.eps}
\includegraphics[width=5.0truein]{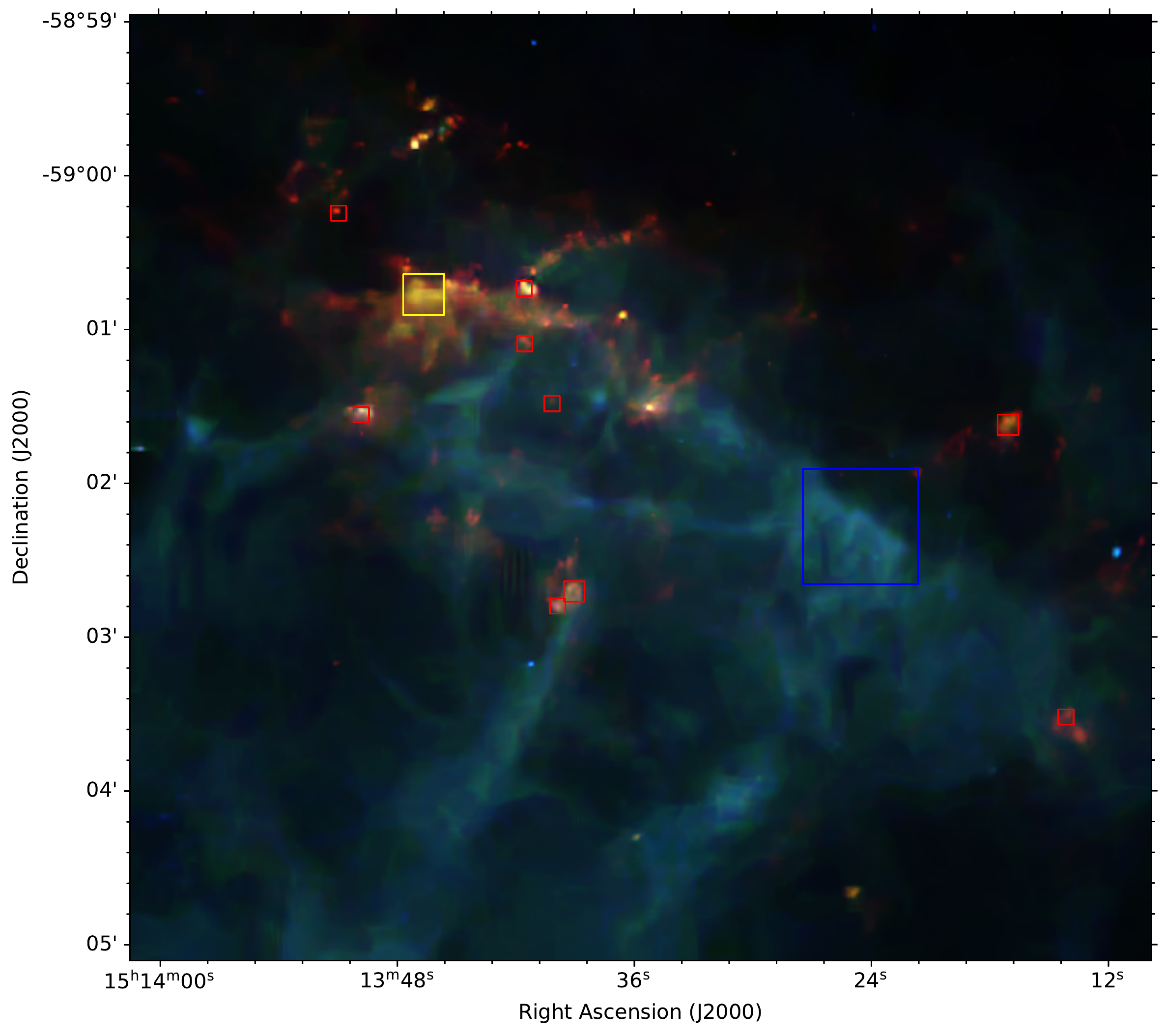}
%\plotone{rcw89colorfigure.eps}
%\includegraphics[width=3truein]{rcw89colorfigure.eps}
\caption{{\sl Chandra} image of RCW 89 in 2018 (red: 0.5--2 keV;
  green: 2--3 keV; blue: 3--6.5 keV). Soft, thermally emitting compact
  ejecta knots are in sharp contrast to diffuse and hard X-rays of nonthermal
  origin. Boxes enclose knots and filaments selected for  
  PM measurements (see their blowups in Figure~\ref{threeepochfig}).}
\label{figcolor}
\vspace{-0.1truein}
\end{center}
\end{figure*}

{\sl Chandra} observations \citep{gaensler02} have shown that RCW
89 was bright in X-rays, with many small bright knots of emission not
well correlated with optical knots, but in some cases coincident with
radio knots \citep{gaensler99}.
\cite{gaensler99} argued convincingly that the MSH 15$-5{\it 2}$/RCW
89 complex is in fact a single SNR, and that its large
size and young age could be understood as the result of a
stripped-envelope supernova (SN Ibc), a suggestion seconded by
\cite{chevalier05}.  A Sedov blast wave could reach the required
radius in 1700 yr for an upstream density of $\sim 0.04$ cm$^{-3}$,
perhaps understandable as the stellar-wind bubble blown by the
progenitor.  In that case, the total swept-up mass is $\sim 9 M_\odot$
and the current shock velocity is $0.4R/t \sim 3000$ km s$^{-1}$.

\cite{yatsu05} showed that
the X-ray knots have individual, differing thermal spectra, based on
the original 2000 20 ks {\sl Chandra} observation.
In an unpublished PhD thesis, \cite{yatsu08} attempted to measure proper
motions (PMs) between two
\cha\ images from 2000 and 2004, with some success, finding velocities
of some knots of order 4000 km s$^{-1}$, in contrast with the low
velocities of optical knots, but consistent with the young age of
\src.  However, the far off-axis position of RCW 89 for one of the
observations resulted in large uncertainties.

These contradictory findings raise various important questions. Where
is the blast wave?  What are its properties? Can we find and identify
SN ejecta, and distinguish them from shocked ambient medium?  What is
the nature of the interaction of the PWN with ejecta?  Is further
particle acceleration to TeV energies occurring, as is observed in all
SNRs as young as \src?  To explore these questions, we obtained a
185 ks observation of \src\ in 2017 and 2018, as described below.

\section{Observations}
\label{obssec}

\src\ was observed by {\sl Chandra} in 2004 December for 29 ks, and
2008 June for 59 ks. Deep (184 ks effective exposure time) third-epoch
observations of RCW 89 include four individual pointings between 2017
December 27 and 2018 February 13 and a final fifth pointing in 2018
April 8--9 (for brevity, we hereafter refer to these five pointings as the 2018 
observations). In all observations, \src\ was placed on the
Advanced CCD Imaging Spectrometer (ACIS) S3 chip, with Very Faint mode used to
reduce the particle background. No significant particle flares were
found.

We aligned individual third-epoch pointings using bright and
compact emission knots (see Figure~\ref{figcolor}). The inter-epoch alignment
was done by matching positions of several point sources near the {\sl Chandra}
optical axis. We jointly fit them with 2D Gaussians to arrive at best estimates of
image shifts between the epochs, with statistical $1\sigma$ 
errors in alignment not exceeding $0.1$ ACIS pixels in R.A.~and decl.~(ACIS pixel
size is
$0\farcs492 \times 0\farcs492$). These errors translate into 
PM errors in R.A.~and decl.~below 5 mas yr$^{-1}$,
given the (exposure-weighted) time baselines of $13.09$ ($9.62)$ yr
between the 2004 (2008) and 2018 observations.

\begin{figure*}[ht!]
\begin{center}
%  \epsscale{1.0}
%  \plotone{f2.eps}
  \includegraphics[width=5.5truein]{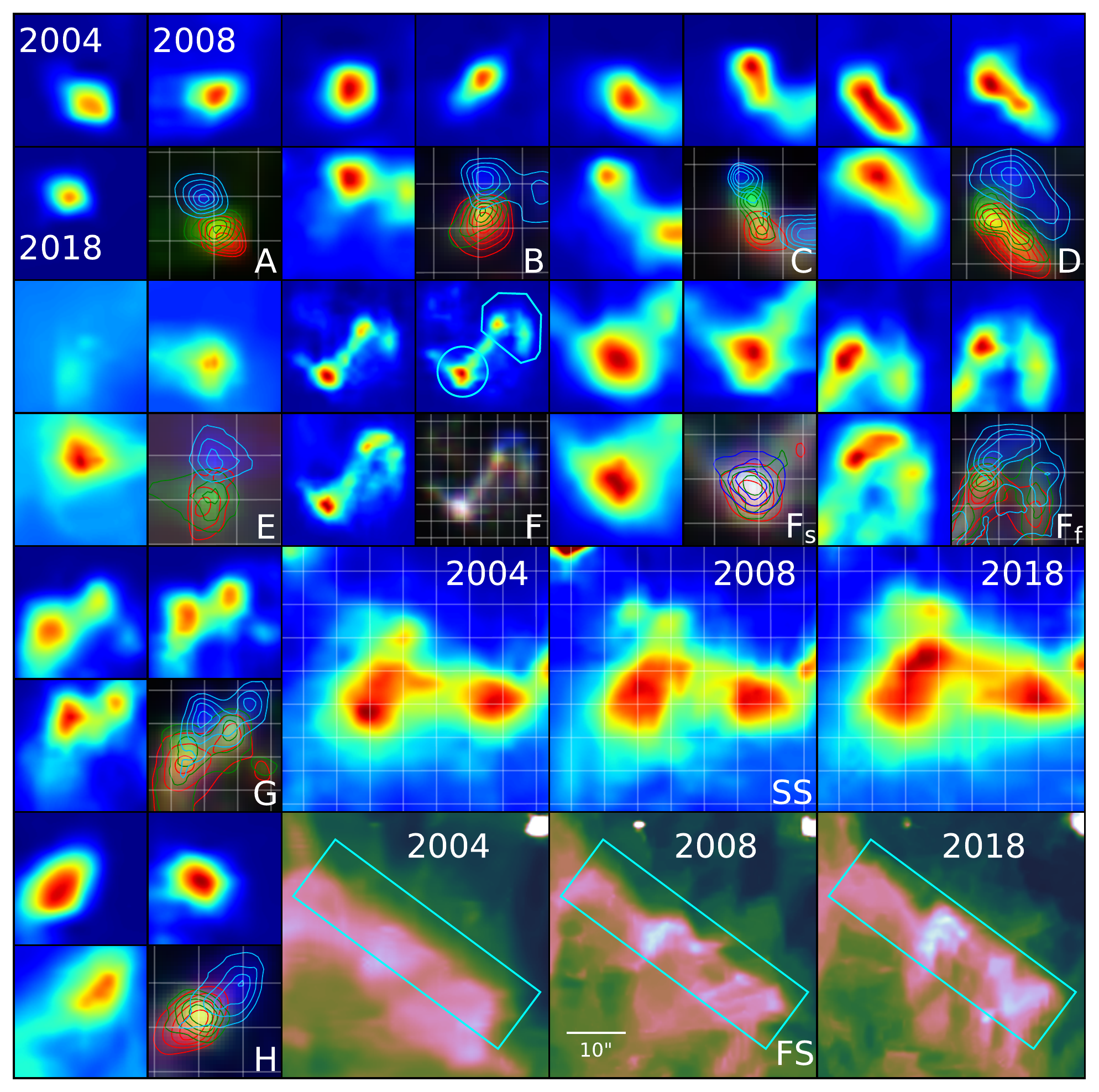}
%  \plotone{rcw89multifig.eps}
%\includegraphics[width=3truein]{g02fig.eps}
  \caption{Selected emission knots and filaments in 2004, 2008, and 2018. Their
    large motions are apparent. RGB images (bottom right of each knot's panels,
    except for regions SS and FS), with contours and coordinate
    grids drawn over, show them in more detail
    (red--2004, green--2008, blue--2018). Knots F$_s$ (bottom left) and F$_f$
    (top right) are within regions drawn over the 2008 image in panel F. Region
    FS (in cyan) is where we measured motion of the fast X-ray synchrotron
    emitting blast wave. Each coordinate grid cell is
    $2\arcsec \times 2\arcsec$ in size.}
  \label{threeepochfig}
\end{center}
\vspace{-0.1truein}
\end{figure*}

After alignment, we extracted images and spectra from event files. The
individual 2018 event files were merged together prior to image
extraction, but spectra were extracted separately from each individual event
file. We then summed them together and averaged their spectral and ancillary
responses by weighting them by individual exposure times.

\section{Expansion}

\begin{figure*}[ht!]
\begin{center}
%  \epsscale{1.0}
%  \plotone{f3.eps}
  \includegraphics[width=5.0truein]{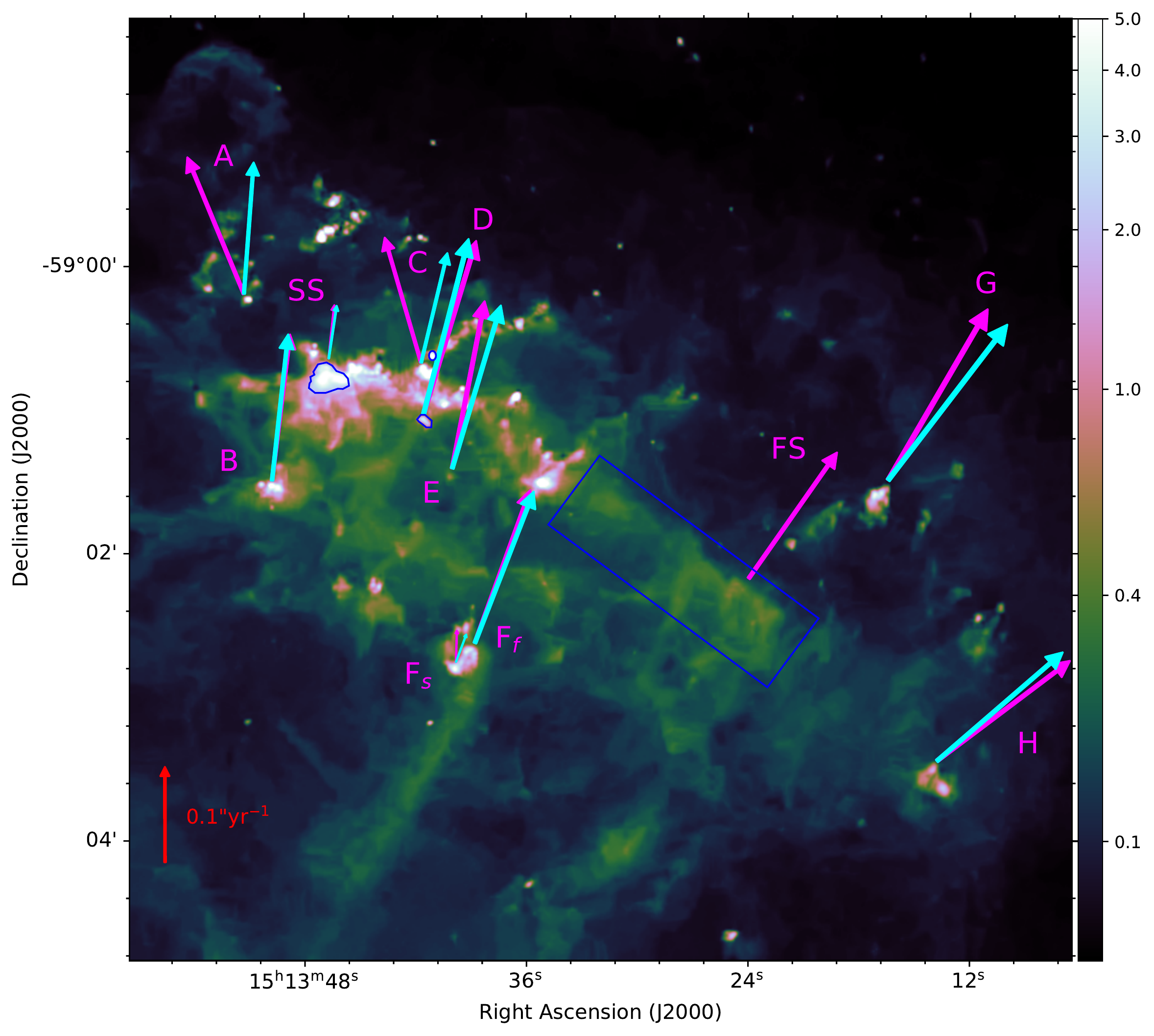}
%  \plotone{rcw89motionsbigfig.eps}
%\includegraphics[width=3truein]{g02fig.eps}
  \caption{
    PMs (magenta arrows, not to scale), and their radial components with
    respect to the pulsar (cyan arrows), drawn over the broadband (0.5--6.5 keV)
    {\sl Chandra} image. The scaling is shown by red arrow pointing radially away
    from the pulsar.
    The four regions (in blue) are where we extracted
    spectra shown in Figure~\ref{figspectra}. The scale is in counts per
    $0 \farcs 192 \times 0 \farcs 192$ image pixel.
  }
  \label{motionsfig}
\end{center}
\vspace{-0.1truein}
\end{figure*}

Motions of most compact emission knots can be readily discerned by
eye: see Figure~\ref{threeepochfig} for a number of selected knots,
and Figures~\ref{figcolor} and \ref{motionsfig} for their location within RCW 89. Additionally,
morphologies of many knots have changed over time 
(knot F$_f$ is an extreme example).  Several knots have either
brightened, including knot A,
or faded (e.g., knots B and H), by up
to a factor of several for both fading and brightening knots. Entirely
new ejecta knots appeared in 2018, including new knots adjacent to
knot B and a bright knot northwest of knot C at R.A.
$= 15^{\rm h}13^{\rm m}41\fs 1$ and $\delta = -59\arcdeg 00\arcmin 38\arcsec$.

PMs were first independently measured on each of the two
(13.1 and 9.6 yr) long time baselines, and then combined (except for
region FS; see below). After extracting broadband (0.5--6.5 keV)
images, $359\arcsec \times 359\arcsec$ in size and with $0\farcs175
\times 0\farcs175$ image pixel, we used the iterative
variance-stabilization method of \citet{azzarifoi16} to smooth them
(parts of these smoothed images are shown in
Figure~\ref{threeepochfig}). Except for knots E and H, we then used a
smoothed image from one epoch and an unsmoothed image from another
epoch to measure motions, modeled as simple image shifts but also
allowing for brightening or fading through an overall change in the
mean surface brightness between the epochs. Poisson statistics were
assumed for unsmoothed images.  A maximum likelihood method was employed,
with smoothed images used as models, and with unsmoothed data taken
only from small regions encompassing measured knots.
On each of the time baselines,we computed a signal-to-noise
ratio-weighted PM from measurements
based on two smoothed and unsmoothed image pairs, with unsmoothed data
and smoothed models alternately taken from each of the two
epochs. Errors were added in quadrature, so the counting noise at
each epoch was taken into account. But for faint and compact knots,
smoothing of images results in a significant loss
in precision. Knot E is the faintest and
most compact among the knots shown in Figure~\ref{threeepochfig},
while knot H 
has faded considerably since 2004 and 2008. So for these two
knots, we just fit a 2D Gaussian (+ constant background) to unsmoothed
data at each epoch \citep[again using C-statistics;][]{cash79}, then
calculated shifts of Gaussian peaks between the epochs and converted
them into PMs.

Results of our PM
measurements can be found in columns (2) and (4) in Table~\ref{rates}. Only
statistical errors are listed; neither errors
in alignment or (unknown but likely substantial) systematic errors are
taken into account. The final 
results in columns (3) and (5) are obtained by combining PMs 
(after adding in quadratures the 5 mas yr$^{-1}$ alignment error in
each individual measurement). We also list in Table~\ref{rates} radial and
tangential PM components with respect to the pulsar, $\mu_r$ and
$\mu_t$, angular
distances of knots from the pulsar, (radial) expansion rates, deceleration
parameters $m$ (products of expansion rates and the estimated remnant's age),
and radial and tangential velocities $v_r$ and $v_t$.

Motion of a low surface brightness filament within region FS can also be
discerned by eye (Figure~\ref{threeepochfig}), but its measurement 
requires use of a different, more advanced technique.
Presumably, this filament is a fast blast wave moving radially away from the
pulsar. In order to measure its radial motion, we used a Bayesian method that
we had employed previously in our study of 
the very young remnant G330.2+1.0 \citep{borkowski18}. Briefly, we used the
smoothed 2018 image (shown at the right-hand corner of
Figure~\ref{threeepochfig}) as a model that was allowed
to expand or shrink, and also vary in its mean surface brightness, with the
expansion centered on the pulsar. Unsmoothed
2004 and 2008 data were jointly fit by this model, assuming a constant
expansion rate across all epochs. In order to verify this assumption, we also
measured expansion rates based only on two-epoch baselines. The mean expansion
rate is $0.0329_{-0.0048}^{+0.0047}\%$ yr$^{-1}$ for the 2008--2018 baseline, and
$0.0494_{-0.0059}^{+0.0057}\%$ yr$^{-1}$ for the 2004--2018 baseline (errors are
$68\%$ credible intervals). These rates are in reasonable agreement with the
rate of $0.0374_{-0.0044}^{+0.0044}\%$ yr$^{-1}$ (see column (9) of 
Table~\ref{rates}) based on the joint fit using data from all three
epochs.

\addtolength{\tabcolsep}{-3pt}
\begin{deluxetable*}{lllllllclcrr}
%\begin{deluxetable*}{lll@{\hspace{15pt}}llllclcrr}
%\rotate
%\tabletypesize{\scriptsize}
%\tabletypesize{\footnotesize}
%\tabletypesize{\small} 
\tablecolumns{12}
\tablecaption{PMs and Expansion Rates\label{rates}}
\tablehead{
  \colhead{Region} & \multicolumn{2}{c}{$\mu_{\alpha}\cos \delta $\tablenotemark{a}} & \multicolumn{2}{c}{$\mu_{\delta}$\tablenotemark{a}} & \multicolumn{1}{c}{$\mu_r$\tablenotemark{b}} & \multicolumn{1}{c}{$\mu_t$\tablenotemark{c}} & Distance &  \multicolumn{1}{c}{Expansion Rate} & $m$\tablenotemark{d} & \multicolumn{1}{c}{$v_r$\tablenotemark{e}} & \multicolumn{1}{c}{$v_t$\tablenotemark{e}} \\
%\cline{2-3} \cline{4-5}
& \multicolumn{2}{c}{(\arcsec yr$^{-1}$)} & \multicolumn{2}{c}{(\arcsec yr$^{-1}$)} & \multicolumn{1}{c}{(\arcsec yr$^{-1}$)} & \multicolumn{1}{c}{(\arcsec yr$^{-1}$)} & (\arcsec) & \multicolumn{1}{c}{(\%~yr$^{-1}$)} &  & \multicolumn{1}{c}{(km s$^{-1}$)} & \multicolumn{1}{c}{(km s$^{-1}$)} }
\decimalcolnumbers
\startdata
A     & $_{\phantom{-} 0.058(8)}^{\phantom{-} 0.059(4)}$  & \phs $0.059(5)$ & $_{0.156(7)}^{0.136(4)}$ & $0.143(5)$ & $0.139(5)$ & $-0.070(5)$ & $476$ & $0.0291 \pm 0.0011$ & $0.50$ & $3420 \pm 120$ & $-1720 \pm 130$\\
B     & $_{-0.022(4)}^{-0.017(3)}$ & $-0.019(4)$ & $_{0.157(4)}^{0.149(3)}$ & $0.153(4)$ & $0.154(4)$ & \phs $0.002(4)$   & $400$ & $0.0384 \pm 0.0011$ & $0.66$ & $3790 \pm 110$ & \phs $60 \pm 110$\\
C     & $_{\phantom{-} 0.037(3)}^{\phantom{-} 0.041(4)}$  & \phs $0.038(4)$ & $_{0.111(3)}^{0.152(3)}$ & $0.131(4)$ & $0.118(4)$ & $-0.068(4)$ & $459$ & $0.0258 \pm 0.0009$ & $0.44$ & $2920 \pm 100$ & $-1670 \pm 110$\\
D     & $_{-0.068(9)}^{-0.047(6)}$ & $-0.055(6)$ & $_{0.179(8)}^{0.182(6)}$ & $0.181(6)$ & $0.189(5)$ & \phs $0.008(7)$ & $439$ & $0.0431 \pm 0.0012$ & $0.74$ & $4660 \pm 130$ & \phs $200 \pm 160$\\
E     & $_{-0.030(20)}^{-0.034(22)}$ & $-0.034(17)$ & $_{0.195(20)}^{0.137(28)}$ & $0.175(15)$ & $0.178(17)$ & $-0.018(16)$ & $418$ & $0.0425 \pm 0.0040$ & $0.73$ & $4380 \pm 380$ & $-440 \pm 420$\\
F$_s$ & $_{\phantom{-} 0.004(8)}^{-0.004(6)}$ & $-0.001(6)$ & $_{0.031(7)}^{0.034(7)}$ & $0.033(6)$ & $0.031(6)$ & $-0.011(6)$ & $343$ & $0.0090 \pm 0.0016$ & $0.15$ & $760 \pm 140$ & $-270 \pm 150$\\
F$_f$ & $_{-0.051(8)}^{-0.062(7)}$ & $-0.058(6)$ & $_{0.149(8)}^{0.172(8)}$ & $0.160(7)$ & $0.170(7)$ & $-0.005(6)$ & $351$ & $0.0485 \pm 0.0020$ & $0.83$ & $4190 \pm 170$ & $-130 \pm 140$\\
G     & $_{-0.106(6)}^{-0.102(5)}$ & $-0.104(6)$ & $_{0.178(6)}^{0.180(6)}$ & $0.179(5)$ & $0.205(6)$ & $-0.026(5)$ & $494$ & $0.0415 \pm 0.0011$ & $0.71$ & $5060 \pm 120$ & $-650 \pm 140$\\
H     & $_{-0.145(12)}^{-0.136(9)}$ & $-0.139(8)$ & $_{0.105(12)}^{0.105(9)}$ & $0.105(8)$ & $0.174(9)$ & \phs $0.011(7)$ & $424$ & $0.0410 \pm 0.0022$ & $0.70$ & $4290 \pm 230$ & $280 \pm 170$\\
SS    & $_{-0.001(8)}^{-0.010(6)}$ & $-0.006(6)$ & $_{0.057(5)}^{0.054(5)}$ & $0.056(5)$ & $0.056(5)$ & $-0.002(6)$ & $448$ & $0.0124 \pm 0.0011$ & $0.21$ & $1380 \pm 120$ & $-60 \pm 150$\\
FS    &  \phs \nodata & \phs \nodata & \nodata & \nodata & $0.161(19)$ & \phs \nodata & $430$ & $0.0374 \pm 0.0044$ & $0.64$ & $3970 \pm 460$ & \nodata \phs \\ 
\enddata
\tablecomments{Errors are $1\sigma$. PM errors (in mas yr$^{-1}$) are listed in parentheses.}
\tablenotetext{a}{Left column: PMs based on the 2004--2018 (top) and 2008--2018 (bottom) time baselines. Right column: combined PM.}
\tablenotetext{b}{Radial PM.}
\tablenotetext{c}{Tangential PM, positive (negative) for clockwise (counter-clockwise) motion.}
\tablenotetext{d}{Deceleration parameter \citep[assuming remnant's age is equal
    to the pulsar's characteristic age of $1710$ yr;][]{livkas11}.}
\tablenotetext{e}{Velocity is $v_{5.2}d_{5.2}$, with $d_{5.2}$ the distance in units of $5.2$ kpc.}
\vspace{-0.2truein}
\end{deluxetable*}

Our measured PMs are shown Figure \ref{motionsfig}, with
their radial components with respect to the pulsar. The fastest
($\mu_r = 205 \pm 6$ mas yr$^{-1}$, $v_r=5060 \pm 120 d_{5.2}$ km
s$^{-1}$) knot G is also the most distant of our knots from the
pulsar. Its motion is nearly radial, but there is a statistically
significant ($v_t=-650 \pm 140 d_{5.2}$ km s$^{-1}$) tangential
component. Its deceleration parameter $m$ of $0.71$ is much less than
unity.  Its complex morphology has changed somewhat since 2004,
suggesting the presence of significant internal motions and/or
brightness variations. Other very fast-moving ($v_r > 4000 d_{5.2}$ km
s$^{-1}$) knots D, E, F$_f$, and H are moving radially away from the
pulsar, but with $m$ still less than $1$. Knot F$_f$ that is located
closer to the pulsar might be somewhat less ($m=0.83$)
decelerated. Another radially moving and rapidly varying knot, knot B,
appears to be somewhat slower ($\mu_r = 154 \pm 4$ mas
yr$^{-1}$, $v_r = 3790 \pm 110 d_{5.2}$ km s$^{-1}$) and more
($m=0.66$) decelerated than other radially moving fast knots. Finally,
still slower and more decelerated knots A and C exhibit strongly
nonradial motions. We caution here that for rapidly brightening (or
fading) knots such as knot A our measured motions might not
necessarily reflect the bulk ejecta motions. Nevertheless, nonradial
motions, brightening and fading, and changes in knot morphologies are
all consistent with the fast-moving, dense ejecta clumps being rapidly
decelerated after a sudden encounter with a dense
ambient medium north of the pulsar.

The closest knot to the pulsar whose motion we measured, knot F$_s$,
stands apart from other compact knots because of its slow ($\mu_r = 31
\pm 6$ mas yr$^{-1}$, $v_r = 760 \pm 140 d_{5.2}$ km s$^{-1}$)
motion. This is most likely a particularly dense cloud in the ambient
medium that has been shocked and accelerated first by the SN blast
wave and then perhaps accelerated even more by subsequent collision
with the SN ejecta.
An X-ray- and radio-bright diffuse filament within region SS also moves
relatively slowly, with $\mu_r = 56 \pm 5$ mas yr$^{-1}$
($v_r = 1380 \pm 120 d_{5.2}$ km s$^{-1}$). This must be a strongly decelerated
blast wave that has encountered a much denser ambient medium than 
the much faster ($v_r = 4000 \pm 500 d_{5.2}$ km s$^{-1}$)
blast wave seen within region FS.

\section{Spectroscopy}

\begin{figure*}
\begin{center}
\hspace{0.5truein}
  %\plotone{rcw89colorfigure.eps}
%\includegraphics[angle=270,width=7.0truein]{f4.eps}
\includegraphics[width=6.0truein]{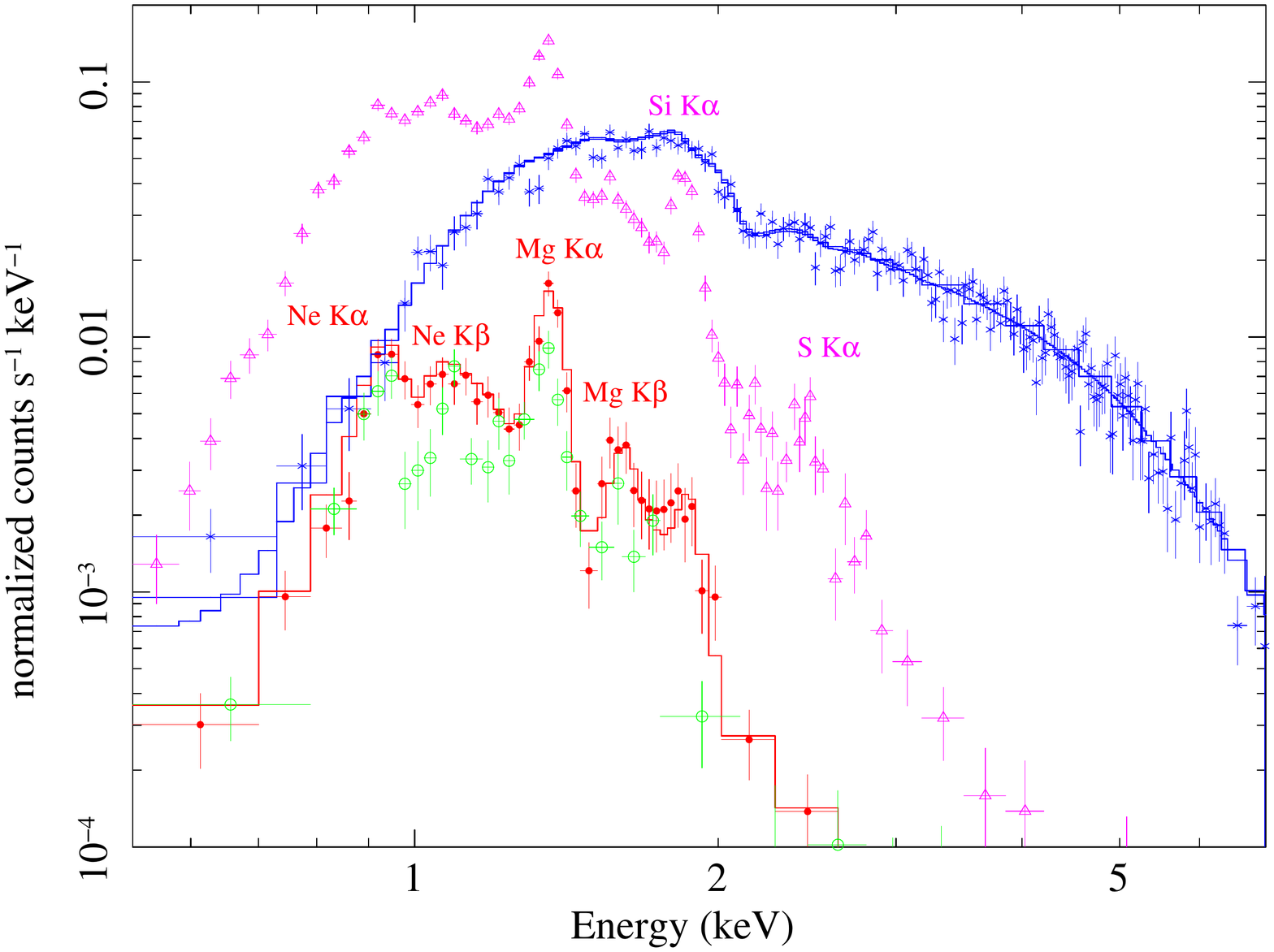}
\caption{Spectra extracted from the regions shown in Figure~\ref{motionsfig}.
  Hard nonthermal spectrum of the fast blast wave (blue stars) contrasts
  sharply with soft thermal spectra of the slowly moving region SS (magenta
  triangles), the fast-moving knot D (open circles in green), and a bright knot
  near knot C that was not seen prior to the 2018 observations (filled red
  circles).
  Spectral models (see the text for more information)
  are also shown for the blast wave (blue line) and one of the ejecta knots 
  (red line).}
\label{figspectra}
\end{center}
%\vspace{-0.2truein}
\end{figure*}

We fit the data with several models from XSPEC v12.10
\citep{arnaud96}, using C-statistics \citep{cash79} and abundances
from \cite{grsa98}.
Backgrounds were modeled
(not subtracted) as required for use with C-statistics.

\subsection{Knots}

Spectra of fast-moving ejecta knots are dominated by very strong lines of
He-like Ne and Mg ions, as shown in Figure~\ref{figspectra} for knot D. These
lines are also strong in knots that became visible only in 2018. We modeled
the spectrum of the brightest of these new knots (see Figure~\ref{figspectra})
by an absorbed plane-shock model with pure heavy-element ejecta containing
only O, F, Ne, Na, Mg, Al, and Si (with odd-$Z$ element abundances set to solar
with respect to O). We obtain a column density
$N_{\rm H} = 1.32 (1.22, 1.43) \times 10^{22}$ cm$^{-2}$, electron temperature
$0.57 (0.37, 1.01)$ keV, ionization age
$\tau = 1.03 (0.38, 1.87) \times 10^{11}$ cm$^{-3}$ s, and blueshift of
$3100 (2500, 3800)$ km s$^{-1}$ (errors are 90\%\ confidence intervals). 
Elemental relative abundances with respect
to solar are $[{\rm Ne}/{\rm O}] = 1.17 (0.90, 1.54)$,
$[{\rm Mg}/{\rm O}] = 0.76 (0.58, 1.01)$, and
$[{\rm Si}/{\rm O}] = 0.19 (0.11, 0.27)$, indicating that these ejecta are
mostly the products of hydrostatic burning.

Lines of Ne and Mg are much weaker in the spectrum of the slowly moving region
SS than in the ejecta knots (Figure~\ref{figspectra}). Thermal continuum is
clearly visible at higher energies, together with the Si and S K$\alpha$
lines. The Si/Mg K$\alpha$ line ratio is much higher than in the
ejecta knots. These are all signatures of normal (cosmic) abundance gas
expected in the shocked ambient medium that must be present.

\subsection{Blast Wave}

We identify the clear edge structure visible in Figure~\ref{figcolor} as the
blast wave, though projection effects may allow some shocked material to
appear farther from the pulsar.  The spectrum of the region shown in 
Figure~\ref{motionsfig} can be well described either by a power law with column
density $N_{\rm H} = 1.23 (1.16, 1.31) \times 10^{22}$ cm$^{-2}$ and photon
index $\Gamma = 2.33 (2.27, 2.40)$, or an srcut model (synchrotron
radiation from a power-law electron distribution with exponential
cutoff, in which, in the absence of radio data, the mean spectral
index was set to $-0.5$) with $N_{\rm H} = 1.12 (1.05, 1.18) \times 10^{22}$
cm$^{-2}$ and rolloff frequency $\nu_{\rm roll} = 2.54 (2.12, 3.28)
\times 10^{17}$ Hz.  Either the power-law index or the rolloff
frequency values are comparable to those seen in other
X-ray-synchrotron-dominated SNR (XSSNR) blast waves
\citep{reynolds08}. 
We attribute the lack of thermal emission to the same cause as in those
other XSSNRs: very low ambient density, as is also necessary for the
blast wave to have traveled this far in 1710 yr. In
this interpretation, the blast wave has not encountered as
much material as we infer to be present in the brighter, more
eastern regions of RCW 89.

We observe an expansion of the edge at a speed of $(4000 \pm
500)d_{5.2}$ km s$^{-1}$.  This is certainly adequate to accelerate
particles to X-ray emitting energies in less than 1700 yr, and
places \src\ among fewer than 20 SNRs showing X-ray synchrotron
emission from the blast wave.  One puzzle is the apparent absence of a
radio counterpart to the X-ray edge in 6 cm observations with the
Australia Telescope Compact Array (ATCA) that were reported by 
\citet{leung18}, but a relatively flat radio-to-X-ray spectral index
could cause the radio feature to be difficult to distinguish among the
diffuse radio emission that permeates \src.

\section{Results and Discussion}

Here we enumerate our results.

\begin{enumerate}
  \itemsep1pt
  \parskip0pt
  \parsep0pt
\item We find that small structures in \src\ are highly dynamic, with
  large PMs in both radial and transverse directions with
  respect to the pulsar, and substantial brightness variations, both
  increasing and decreasing, over the 14 yr covered by the three
  observations.
\item Expansion velocities for small-scale features range from 
  5000 km s$^{-1}$ for the fastest knots to less than 1000 km s$^{-1}$ for
  one feature.  However, decelerations are large because the mean
  expansion velocity for the outer material is about 13,000 km
  s$^{-1}$.
\item Most knots have spectra clearly indicative of ejecta, in
  particular strong lines from Ne and Mg, material synthesized
  hydrostatically rather than explosively.
  \item We have located a feature we identify with the blast wave,
  expanding at 4000 km s$^{-1}$, with a lineless spectrum we
  attribute to synchrotron radiation.  This feature has no radio
  counterpart.

\end{enumerate}
  
  The rapid brightness changes in small knots require high densities.
  For a knot to appear in 10 yr or less requires an electron
  density of several hundreds, from our fitted ionization timescales
  $n_e t$ of order $10^{11}$ cm$^{-3}$ s.  While knot velocities are
  high by absolute standards, the knots have been substantially
  decelerated in the 1700 yr since the explosion.  This
  deceleration must have occurred fairly recently.

  In general, we have a picture of a blast wave from a
  stripped-envelope SN racing through a low-density wind-blown
  bubble, and impacting the bubble wall initially in the north.  The
  SN ejecta are mainly diffuse, but some very dense clumps are
  visible as they impact the wall and shocks are driven into them.
  The blast wave appears still to be encountering relatively
  low-density material, and no thermal component of the emission is
  detectable.

  Ejecta with velocities this high have been examined in detail in Cas
  A, the youngest \citep[350 yr old;][]{thor01} core-collapse SNR in
  our Galaxy.
  The densest ejecta in Cas A are optically emitting fast-moving knots
  (FMKs) that have suffered little deceleration so far. Their
  estimated preshock densities are uncertain, but they might be as
  high as 100--300 atoms cm$^{-3}$ \citep{docenko10}. If such knots
  expand freely for 1710 yr, their preshock densities would drop by a
  factor of $\left(1710\ {\rm yr}/350\ {\rm yr}\right)^3 = 120$. When
  rapidly shocked and strongly decelerated, they would emit at X-ray
  (not optical) wavelengths.  After compression by a factor of 4 in a
  strong shock, electron densities would be in the range of 30--90
  cm$^{-3}$ for the shocked plasma ionization state and abundances we
  find in the bright new knot whose spectrum is shown in
  Figure~\ref{figspectra}. This
  is only 4--12 times lower than the density inferred by us for this knot.
  We conclude that the fast-moving ejecta clumps in
  \src\ resemble those seen in Cas A as FMKs in terms of their initial
  velocities and densities. They might have the same, still not
  understood origin, presumably related to stripped-envelope SN
  explosions themselves.

  Various proposals have been floated for the relation between the
  pulsar and PWN and RCW 89.  We find little or no evidence for the
  direct interaction of pulsar-produced phenomena and the X-ray
  emission of RCW 89.  Diffuse emission away from the bright knots has
  a nonthermal spectrum, but its relatively steep spectral index means
  it could either be post-shock emission from shock-accelerated
  electrons, or loss-steepened emission from electrons escaping from
  the PWN.  
   
  The detection of significant X-ray synchrotron emission from the
  blast wave in \src\ makes it the only known SNR showing both a
  PWN and forward-shock nonthermal X-ray emission.

\vspace{5mm}
%  \acknowledgments
We acknowledge support by NASA through {\sl Chandra} General Observer Program 
grant SAO GO7-18068X.

\facilities{CXO}

\end{document}